\documentclass[
aps,
prl,
reprint,
superscriptaddress,
longbibliography,
floatfix
]{revtex4-2}

\usepackage[T1]{fontenc}
\usepackage[utf8]{inputenc}
\usepackage{lmodern}
\usepackage{amsmath,amssymb}
\usepackage{graphicx}
\usepackage{xcolor}
\usepackage{xurl}

\usepackage[
    colorlinks=true,
    linkcolor=blue,
    citecolor=blue,
    urlcolor=blue
]{hyperref}

\setcitestyle{super,sort&compress}

\makeatletter
\def\fnum@figure{\textbf{Figure \thefigure}}
\makeatother

\begin{document}

\title{Quantum Back-Action Expands the Excitonic Hilbert Space in a Soft Polar Semiconductor}

\author{Arnab Ghosh}
\affiliation{Department of Chemistry, McGill University, Montreal, H3A 0B8, Canada}
\author{Patrick Brosseau}
\affiliation{Department of Chemistry, McGill University, Montreal, H3A 0B8, Canada}
\author{Dmitry N. Dirin}
\affiliation{Department of Chemistry and Applied Biosciences, ETH Z\"urich, Switzerland}
\affiliation{Laboratory for Thin Films and Photovoltaics, Empa - Swiss Federal Laboratories for Materials Science and Technology, Switzerland}
\author{Priya Nagpal}
\affiliation{Department of Chemistry, McGill University, Montreal, H3A 0B8, Canada}
\author{Rui Tao}
\affiliation{Department of Chemistry and Applied Biosciences, ETH Z\"urich, Switzerland}
\affiliation{Laboratory for Thin Films and Photovoltaics, Empa - Swiss Federal Laboratories for Materials Science and Technology, Switzerland}
\author{Maksym V. Kovalenko}
\affiliation{Department of Chemistry and Applied Biosciences, ETH Z\"urich, Switzerland}
\affiliation{Laboratory for Thin Films and Photovoltaics, Empa - Swiss Federal Laboratories for Materials Science and Technology, Switzerland}
\author{Patanjali Kambhampati}
\affiliation{Department of Chemistry, McGill University, Montreal, H3A 0B8, Canada}

\begin{abstract}
Electronic excitations in solids are usually described by a hierarchy in
which the Hamiltonian is fixed first and the lattice acts afterward,
renormalizing energies, scattering populations, or degrading coherence.
That picture assumes that the optically accessible excitonic manifold is
already present when light arrives. Here we show that this assumption
fails in a soft polar semiconductor. Using femtosecond coherent
multidimensional spectroscopy on lead-halide perovskite nanocrystals, we
observe quantum back-action between an electronic excitation and a
collective lattice-polarization field expand the excitonic Hilbert space
in real time. The optical pulse first prepares a single excitonic
polarization, \(X_{1}\). A second configuration, \(X_{2}\), appears only
as the polaron field forms, while coherent anti-diagonal coupling
between \(X_{1}\)and \(X_{2}\) develops later. State formation and
coherence formation are therefore resolved as distinct stages of
quasiparticle birth. CdSe quantum dots provide the conventional
reference limit: confined excitons whose diagonal and anti-diagonal
responses are present at time zero and weakly perturbed by a discrete
phonon bath. The perovskite nanocrystals realize the opposite limit.
They behave as finite pieces of a soft polar semiconductor, where strong
coupling to a broad low-frequency spectral density makes the bath an
active quantum field rather than a perturbative reservoir. The resulting
diagonal and anti-diagonal splittings increase with nanocrystal edge
length, opposite to single-particle confinement, and track the growth of
radiative oscillator strength. A dynamical polaron-field model describes
the lattice polarization as an order parameter that expands the
optically accessible manifold and generates time-dependent coherent
coupling. These results establish strong system--bath coupling as a
constructive mechanism in excitonic condensed matter: the environment
does not merely dress or destroy quantum states, but can create the
manifold in which they coherently evolve.
\end{abstract}
\maketitle

\subsection{1. Introduction}

Excitations in solids are rarely bare particles. They are dressed by
polarization fields, scattered by lattice motion, stabilized by
collective distortions, and sometimes transformed into quasiparticles
whose identity cannot be assigned to the electronic system alone
\cite{ref1,ref2,ref3,ref4,ref5,ref6,ref7,ref8,ref9,ref10}. Yet the standard description of quantum-confined
semiconductors usually begins from a fixed hierarchy: confinement and
Coulomb interactions define the excitonic Hamiltonian, while the lattice
supplies corrections---energy renormalization, dephasing, relaxation,
and vibronic structure.

This separation has been remarkably successful for rigid covalent
semiconductors and related nanostructures such as CdSe and its quantum
dot form. In such rigid systems the excitonic structure exists before
the lattice moves and phonons act mainly as perturbative normal modes
\cite{ref11,ref12,ref13,ref14,ref15,ref16,ref17,ref18,ref19,ref20,ref21,ref22,ref23,ref24}. But it also contains a strong assumption: the
excitonic Hilbert space is already present at the moment of optical
excitation. In a soft polar lattice, that assumption can fail. If the
lattice polarization field reorganizes on the same timescale as the
optical response \cite{ref8,ref9,ref25,ref26,ref27,ref28,ref29,ref30,ref31,ref32}, the environment is no
longer merely a bath. It becomes a dynamical field capable of generating
the potential landscape in which excitons exist.

Several lines of work point toward this regime without directly
resolving it. Raman, terahertz, optical Kerr, photoemission, and
ultrafast optical measurements have established that lead-halide
perovskites possess large-amplitude low-frequency lattice motion, strong
Fröhlich coupling, dynamic disorder, and polaronic screening
\cite{ref8,ref9,ref27,ref28,ref29,ref31,ref32,ref33,ref34,ref35}. Coherent multi-dimensional
spectroscopy has shown liquid-like spectral diffusion in perovskite bulk
nanocrystals, identifying a solvation-like electronic response
associated with polaron formation \cite{ref25,ref26,ref27,ref30}. Related
studies have connected the same soft lattice physics to hot-carrier
protection \cite{ref36,ref37,ref38}, defect tolerance
\cite{ref39,ref40,ref41}, biexciton stabilization
\cite{ref42,ref43,ref44,ref45}, cooperative emission as single dot
superradiance \cite{ref46,ref47}, and unusually robust electronic
coherence \cite{ref29}. These observations make clear that the
lattice in perovskites is not a passive perturbation. It participates
actively in the optical excitation and can dominate the subsequent
many-body response.

What remains missing is a direct measurement of whether this
participation merely dresses pre-existing excitons or instead creates
new excitonic states and couplings during the pulse sequence experiment
itself. A polaron being formed is inferred \cite{ref48,ref49,ref50,ref51,ref52,ref53,ref54,ref55,ref56,ref57,ref58} from
energy shifts, linewidth evolution, phonon spectra, or relaxation
dynamics, but these observables do not by themselves reveal whether the
excitonic Hilbert space is static or dynamically generated.

The decisive experiment must separate three processes that are usually
conflated: optical preparation of an exciton, formation of the polaronic
energy landscape, and emergence of coherent coupling within that
landscape. Without such separation, polaron formation remains a
renormalization problem rather than a state-formation problem, and the
central question remains unresolved: can a collective lattice
polarization field generate an emergent excitonic manifold in real time?
The central result is that, in a soft polar semiconductor, the excitonic
Hamiltonian is not simply uncovered by the optical pulse; it is
assembled by the ensuing lattice response.

Femtosecond coherent multidimensional spectroscopy resolves this
assembly directly. The field first prepares \(X_{1}\), an electronic
polarization in a lattice that has not yet reorganized. The polar
lattice then reacts to the excitation, develops a collective
polarization field, and feeds back on the electronic system. Only then
does \(X_{2}\) enter the optically accessible manifold. Coherent
anti-diagonal coupling follows later, after the manifold has already
formed. The experiment therefore watches a Hamiltonian acquire new
states and then acquire coherence between them. This is not the
conventional physics of confined excitons dressed by phonons. CdSe
realizes that limit: the excitonic manifold is present at time zero and
the bath acts perturbatively. The perovskite response lies in the
opposite regime. Strong coupling to a broad low-frequency lattice
spectral density produces quantum back-action, and that back-action
expands the excitonic Hilbert space in real time.

\subsection{2. Experimental observation of an evolving excitonic manifold}

The spectra reveal an energy scale that does not belong to confinement.
The diagonal and anti-diagonal splittings are born at different times
and measure different parts of the evolving Hamiltonian. The diagonal
splitting appears when the lattice polarization has reorganized enough
to define a new exciton-polaron landscape. The anti-diagonal splitting
appears only later, when phase-coherent coupling develops inside that
landscape. Their finite-size dependence is therefore not the usual
nanocrystal scaling of electronic levels. It is the opposite limit: the
larger the polarizable crystal volume, the larger the collective lattice
field that can participate in the excitation. The relevant object is not
a particle in a box, but an extended polarization cloud in a soft ionic
semiconductor. The finite crystal merely makes this bulk-like field
visible. The energy scale is not quantum-dot fine structure, not a
vibronic replica, and not an accidental shell splitting. It is the
matrix element of a lattice-generated many-body state.

We describe this regime with a dynamical polaron-field picture. The
optical field does not project onto a completed excitonic spectrum. It
launches an electronic polarization into a lattice that has not yet
chosen its configuration. The lattice then moves. Its collective
polarization grows, and the induced field acts back on the electronic
excitation. This back-action changes the accessible Hilbert space:
\(X_{1}\) is no longer the whole response, and \(X_{2}\) becomes part of
the exciton--polaron manifold. The Hamiltonian is therefore
time-dependent in the strongest sense. Its basis changes as the
quasiparticle is born. Once the manifold exists, the same polar field
can generate off-diagonal coupling, but only after the phase relation
between \(X_{1}\) and \(X_{2}\) has developed. This is why the AD
splitting lags the diagonal splitting. The bath first creates the
landscape; coherence comes later.

This is the central result. In weakly coupled excitonic systems, the
environment is appended to an electronic Hamiltonian as a perturbation.
It shifts lines, broadens lines, and removes coherence. In the soft
perovskite lattice, that separation breaks down. The lattice is not
outside the quantum system. It is the slow, polar, strongly coupled
field through which the exciton becomes a many-body object. The
environment is promoted from bath to dynamical actor: it does not merely
dress a state that already exists, but supplies the collective
coordinate that expands the excitonic Hilbert space and then couples the
states within it. This is quantum back-action observed as spectroscopy.

\textbf{Figure 1} defines the conceptual experiment. In the conventional limit,
the optical field interrogates an excitonic spectrum that is already
present. The electronic Hamiltonian is fixed before the pulse arrives;
the lattice then acts as a perturbing reservoir that shifts energies,
broadens transitions, or modulates phase. CdSe realizes this limit:
confined excitons are defined by the static nanocrystal Hamiltonian and
weakly dressed by a discrete phonon bath. The soft polar semiconductor
studied here is different.

\begin{figure*}[t]
\centering
\includegraphics[width=0.75\textwidth]{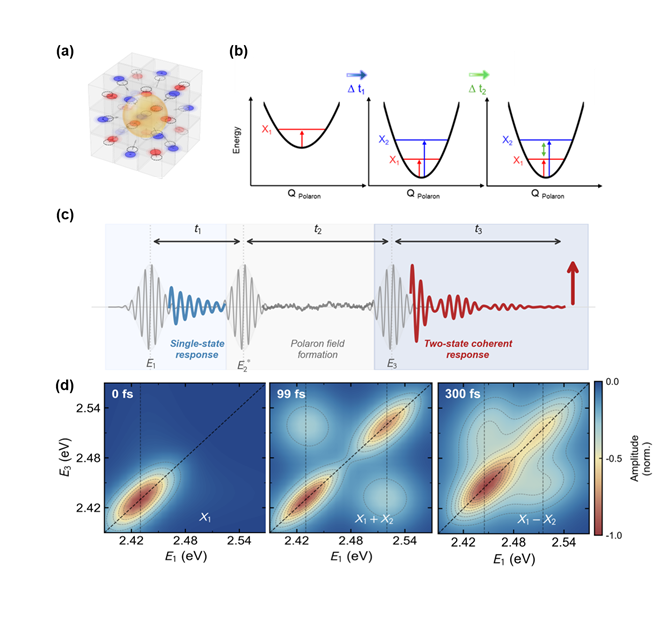}
\caption{Dynamical formation of a coupled exciton--polaron manifold. (a) Photoexcitation in a soft perovskite quantum dot creates an exciton coupled to a collective lattice-polarization field. The relevant quasiparticle is therefore a self-consistent exciton--polaron rather than a bare exciton in a rigid lattice. (b) Minimal dynamical level scheme. The optical field initially prepares \(X_{1}\). During the population time \(t_{2}\), lattice reorganization stabilizes \(X_{2}\) within the polaron-renormalized manifold. The diagonal splitting \(\Delta_{1}\) reports the formation of distinct excitonic configurations, whereas the anti-diagonal splitting \(\Delta_{2}\) reports the later emergence of coherent \(X_{1}\)-\(X_{2}\) coupling. (c) CMDS pulse sequence. The first interactions prepare the optical response, \(t_{2}\)allows the polaron field to form, and the final interaction probes the reorganized exciton--polaron manifold. (d) Temporal hierarchy tested in the experiment: prompt optical preparation, delayed state formation, and still later coherence onset. (e) Simulated 2D spectra illustrating the expected evolution from an initially single-state response, to a two-state exciton--polaron manifold, to a coherently coupled manifold with resolved diagonal and anti-diagonal structure.}
\label{figure1}
\end{figure*}

The optical field does not merely project onto pre-existing excitonic
eigenstates. It launches an electronic polarization into a lattice whose
collective polarization has not yet formed. The lattice responds to that
excitation, develops a long-range polar field, and feeds back on the
electronic system. This quantum back-action expands the optically
accessible excitonic Hilbert space: \(X_{1}\) is prepared first,
\(X_{2}\) becomes visible only after the polaron field develops, and
coherent \(X_{1}\)-\(X_{2}\) coupling appears later still. The central
object is therefore not a confined exciton dressed perturbatively by
phonons, but a dynamical exciton--polaron field in which electronic
polarization and lattice polarization form one self-consistent many-body
state.

\textbf{Figure 1a} gives the microscopic origin of this failure of the
static-Hamiltonian picture. A photoexcited electron--hole pair in a
lead-halide perovskite is embedded in a soft ionic framework
\cite{ref59,ref60,ref61,ref62}, not in a rigid covalent cage as in covalent QD
such as CdSe \cite{ref63}. The soft perovskite lattice is not a
passive background. It polarizes, distorts, and redistributes charge on
the same femtosecond timescale on which the optical excitation is being
formed.

The relevant coordinate is therefore not an external bath coordinate
appended to an electronic Hamiltonian, but the collective polarization
field that stabilizes the quasiparticle itself. In this regime,
``exciton'' and ``environment'' cannot be separated into a fast
electronic system and a slow spectator lattice. The excitation drives
the lattice; the lattice field acts back on the excitation; and the
resulting self-consistent field reshapes the states available to the
optical response. The exciton is therefore not first created and then
dressed. It is born as an exciton--polaron, with its spatial extent,
energy landscape, and internal couplings determined by the quantum
back-action of the polar lattice.

\textbf{Figure 1b} translates this physical picture into the minimal dynamical
level scheme. At time zero, the optical field prepares \(X_{1}\). The
second state, \(X_{2}\), is not assumed to be an optically resolved
member of a pre-existing doublet. It becomes visible only after the
collective polarization field has developed enough to stabilize a second
configuration within the exciton--polaron manifold
\cite{ref64,ref65}. This produces the diagonal splitting,
\(\Delta_{1}\). A still later process produces the anti-diagonal
splitting, \(\Delta_{2}\), associated with coherent off-diagonal
coupling between \(X_{1}\) and \(X_{2}\). The distinction is decisive.
\(\Delta_{1}\) is the energy scale of state formation. \(\Delta_{2}\) is
the energy scale of coherence formation. They need not appear together,
need not be equal, and need not have the same dynamics.

\textbf{Figure 1c} shows why coherent multidimensional spectroscopy
\cite{ref25,ref26,ref27,ref28,ref29,ref66,ref67,ref68,ref69} is the required experiment. The first two
interactions prepare the optical polarization; the population time
\(t_{2}\) is the waiting period during which the lattice field forms; the
third interaction probes the Hamiltonian after that reorganization has
occurred. CMDS therefore does not merely measure a time-resolved
spectrum. It watches the spectrum being constructed. At early \(t_{2}\),
the emitted field is the response of the optically prepared
\(X_{1}\) configuration. At later \(t_{2}\), after the polaron field has
reorganized the lattice, the emitted field contains the response of a
larger exciton--polaron Hilbert space.

\textbf{Figure 1d} states the temporal hierarchy tested in the paper: preparation
first, state formation second, coherence formation third. This is the
signature of an evolving Hamiltonian. If \(X_{1}\), \(X_{2}\), and
\(V_{12}\) were already present in the bare quantum dot Hamiltonian, the
diagonal and anti-diagonal structures would appear together within the
instrument response. The proposed sequence is instead causal: the pulse
prepares \(X_{1}\); the lattice polarization generates the
\(X_{1}/X_{2}\) manifold; only after that can coherent
\(X_{1}\)-\(X_{2}\) coupling appear.

\textbf{Figure 1e} gives the spectral grammar used throughout the work. The
early-time spectrum should be \(X_{1}\)-like, with no resolved
anti-diagonal coherent structure. At intermediate \(t_{2}\), the second
excitonic configuration emerges as the polaron landscape forms. At long
\(t_{2}\), the spectrum acquires anti-diagonal structure from the
off-diagonal density-matrix element connecting \(X_{1}\) and \(X_{2}\).
Thus the diagonal direction reads the birth of states, while the
anti-diagonal direction reads the birth of coherence between them. This
separation is the experimental handle on a central claim: in a soft
polar quantum dot, the lattice does not merely dress the excitonic
spectrum. It generates it.

\textbf{Figure 2} shows the experimental realization of the sequence proposed in
Figure 1. CsPbBr\({}_{3}\) and CsPbI\({}_{3}\) perovskite bulk
nanocrystals were synthesized as described previously
\cite{ref62} and in the Supplementary Information (SI),
dispersed under optically dilute conditions, and circulated through a
flow cell during measurement. These nanocrystals are not physically
confined quantum dots since their edge length is up to 18 nm, compared
to the exciton Bohr diameter of 7 nm.

\begin{figure*}[tbp]
\centering
\includegraphics[width=0.95\textwidth]{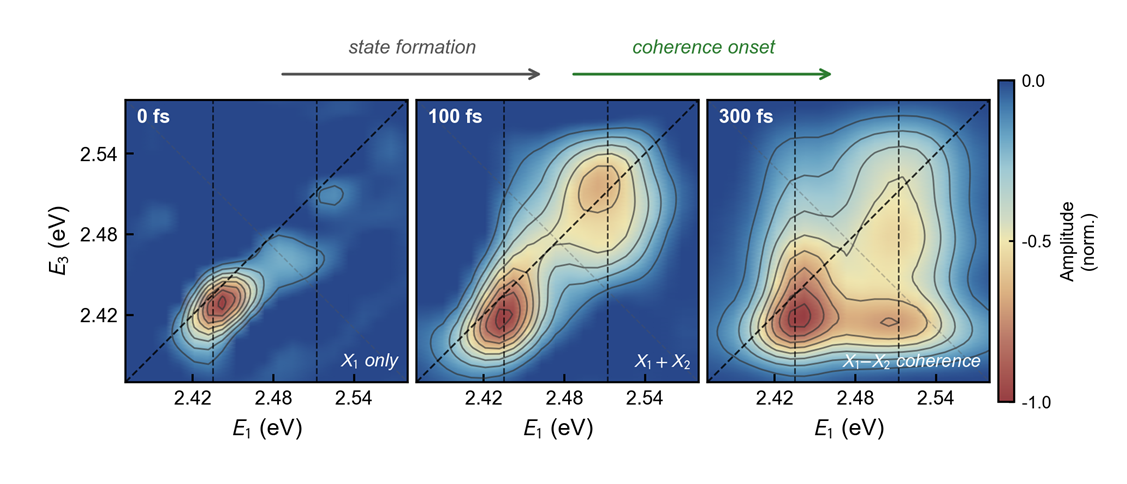}
\caption{Experimental observation of delayed two-state coherence in a perovskite quantum dot. Absorptive coherent multi-dimensional spectra measured at representative population times \(t_{2} = 0\), 100, and 300 fs in CsPbBr\textsubscript{3} bulk nanocrystals of 18 nm edge length. At \(t_{2} = 0\) fs, the response is dominated by the initially prepared \(X_{1}\) transition. By 100 fs, a second excitonic feature appears as the polaron field forms. At long \(t_{2}\), the spectrum develops a resolved anti-diagonal structure assigned to coherent \(X_{1}\)-\(X_{2}\) coupling. The dashed diagonal and vertical guides mark the spectral axes used to distinguish diagonal state splitting from anti-diagonal coherence onset.}
\label{figure 2}
\end{figure*}

Coherent multidimensional spectra were acquired using the same
femtosecond phase-resolved CMDS platform described in the SI and in our
prior works on CMDS method development \cite{ref70,ref71,ref72,ref73} and key
implementations to solve key questions \cite{ref19,ref25,ref26,ref27,ref28,ref29,ref74,ref75}.
Briefly, broadband blue pulses were generated by an Ar-filled
hollow-core fiber pumped by an optical parametric amplifier, and
phase-controlled pulse trains were produced using acousto-optic
programmable dispersive filters. The experiment was performed in a
pump--probe geometry in the rotating frame, with the emitted signal
fields heterodyne detected by spectral interferometry on a CCD. The
absorptive two-dimensional spectra do not reveal a fixed excitonic
manifold immediately after photoexcitation. Instead, the response
evolves in time from a single-state spectrum, to a two-state manifold,
and finally to a coherently coupled two-state response. This temporal
ordering is the first evidence that the excitonic structure is generated
by lattice reorganization rather than inherited from a static
confinement Hamiltonian.

At \(t_{2} = 0\) fs, the spectrum is dominated by a single diagonal
feature assigned to the initially prepared \(X_{1}\) transition. There is
no clearly resolved second state and no developed anti-diagonal
connectivity within the time-zero window. This is the essential starting
condition: the optical field prepares an electronic polarization, not
the full exciton--polaron manifold. By \(t_{2} = 100\) fs, the spectrum
has changed qualitatively. A second feature appears at higher detection
energy, producing a two-state structure. We identify this as state
formation: the lattice polarization field has reorganized sufficiently
to stabilize \(X_{2}\)within the polaron-renormalized energy landscape.
The diagonal separation between the features is therefore not an
ordinary single-particle level spacing, but the spectral signature of an
emergent excitonic manifold.

At \(t_{2} = 300\) fs, the spectrum develops pronounced anti-diagonal
connectivity between the two features. This late-time structure is the
signature of coherent \(X_{1} - X_{2}\) coupling within the already
formed manifold. The ordering is decisive. The two-state manifold
appears first; coherent anti-diagonal coupling appears later. Thus state
formation and coherence formation are experimentally distinct processes.
The dashed diagonal and vertical guides define the analysis geometry
used below: diagonal structure reports the energetic separation of
excitonic configurations, whereas anti-diagonal structure reports
correlated excitation--detection response and therefore coherent
inter-state coupling.

\textbf{Figure 3} reduces the evolving CMDS response to three projections of the
same two-dimensional spectrum \cite{ref66,ref67,ref76,ref77}. The
diagonal projection \(D\) measures the energy separation between
excitonic configurations. The full anti-diagonal projection
\(AD_{splitting}\) measures coherent coupling between them. The local
anti-diagonal projection through the \(X_{1}\) feature, \(AD_{X_{1}}\),
measures the linewidth dynamics and therefore the spectral diffusion
associated with polaron formation. These observables are not redundant
cuts through a static doublet. They isolate different elements of an
evolving exciton--polaron Hamiltonian: the formation of an energy
landscape, the local fluctuation dynamics of that landscape, and the
later development of coherent off-diagonal coupling.

\begin{figure*}[tbp]
\centering
\includegraphics[width=0.70\textwidth]{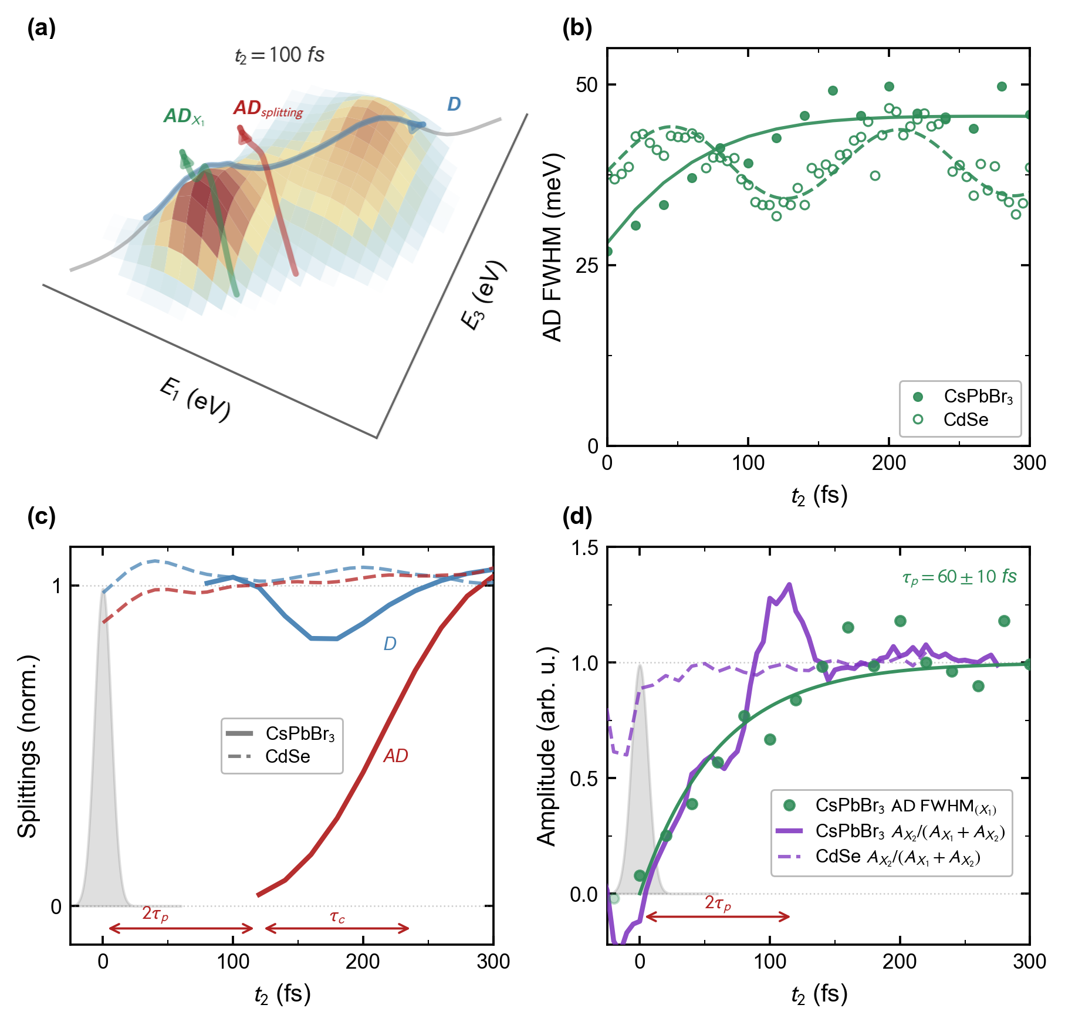}
\caption{CMDS projections separate polaron formation from coherent coupling. (a) Representative \(t_{2} = 100\)fs CMDS response showing the three projections used in the analysis. The diagonal projection \(D\) measures the energy separation of the polaron-generated excitonic configurations. The full anti-diagonal projection \(AD_{splitting}\) measures coherent inter-state coupling. The local anti-diagonal projection \(AD_{X_{1}}\) measures the linewidth of the \(X_{1}\) feature and tracks spectral diffusion during polaron formation. (b) Time dependence of the local \(AD_{X_{1}}\) linewidth. 18 nm edge length CsPbBr\({}_{3}\) shows a monotonic linewidth increase, whereas 3.9 nm diameter CdSe shows oscillatory modulation without comparable buildup, identifying the perovskite response as diffusive lattice reorganization rather than coherent phonon motion. (c) Population-time evolution of the D and AD splittings. In CsPbBr\({}_{3}\), the D splitting appears with polaron formation and is nearly static, while the AD splitting continues to grow from \(\sim 100\) to 300 fs. CdSe shows no delayed AD growth, consistent with a pre-existing excitonic manifold. (d) The AD linewidth dynamics and fractional \(X_{2}\) amplitude give the same polaron-formation time, \(\tau_{p} = 60 \pm 10\) fs, and saturate by \(\sim 200\) fs. Thus the D/AD peak structure appears when the polaron field forms, whereas the later AD splitting reflects delayed \(X_{1}\)-\(X_{2}\) coherence within the already formed exciton--polaron manifold.}
\label{fig:3}
\end{figure*}

\textbf{Figure 3a} defines these projections on a representative CMDS response at
\(t_{2} = 100\ fs\). The \(D\)projection follows correlated excitation
and detection energies and reports the separation of the
polaron-generated excitonic configurations. The
\(AD_{splitting}\) projection cuts across the two-state response and
isolates the coherent inter-state coupling. The local
\(AD_{X_{1}}\) projection instead probes the linewidth of the initially
prepared feature and provides a direct clock for spectral diffusion.
Thus CMDS separates quantities that would be collapsed in a
one-dimensional transient spectrum: state separation, coherence, and
local bath-driven broadening.

\textbf{Figure 3b} shows the linewidth clock. In CsPbBr\({}_{3}\), the
\(AD_{X_{1}}\) linewidth increases monotonically on the
\(\tau_{p}\) timescale, whereas CdSe shows only oscillatory modulation
without comparable growth. The perovskite response is therefore not the
motion of a discrete underdamped phonon superposed on a fixed exciton.
It is a diffusive reorganization of the polar lattice around the
excitation \cite{ref25,ref26,ref27,ref30}.

\textbf{Figure 3c} shows how this lattice clock relates to the \(D\)and
\(AD\) splittings. In CsPbBr\({}_{3}\), the \(D\) splitting appears with
the onset of the polaronic response and then remains nearly static,
indicating that the two-state energy landscape has formed. The
\(AD\) splitting appears only after this manifold exists and then
continues to grow from \(\sim 100\) to \(300\ fs\). Its onset marks the
availability of the \(X_{1}/X_{2}\) manifold; its later growth marks the
maturation of coherent \(X_{1}\)-\(X_{2}\) coupling. CdSe shows no
analogous delayed \(AD_{growth}\), consistent with excitonic states and
couplings that are already present in the static Hamiltonian.

\textbf{Figure 3d} closes the loop by comparing the linewidth dynamics with the
fractional \(X_{2}\) amplitude. Both give the same polaron-formation
time, \(\tau_{p} = 60 \pm 10\ fs\), and saturate by \(\sim 200\ fs\).
The \(D\) and \(AD\) peak structures therefore appear when the polaron
field forms. The continued growth of the \(AD_{splitting}\) after this time
is not simply polaron formation; it is the subsequent growth of the
off-diagonal density-matrix element inside the already formed
exciton--polaron manifold.

\textbf{Figure 4} shows the raw dynamical basis for this interpretation. Rather
than reducing the data immediately to extracted peak separations, the
figure displays the full time-dependent D and AD projections for
CsPbBr\({}_{3}\) and CdSe. \textbf{Figure 4a} shows the diagonal projection of
CsPbBr\({}_{3}\). The split structure appears promptly after the onset
and changes only modestly thereafter, indicating that the energetic
separation of the two excitonic configurations is established rapidly
once the polaron-renormalized landscape forms. \textbf{Figure 4b} shows the
corresponding anti-diagonal projection. This response is qualitatively
different: the AD structure is weak at early time and grows into a
well-defined split response as \(t_{2}\) increases. The AD projection
therefore directly exposes the delayed formation of coherent
\(X_{1} - X_{2}\) coupling.

\textbf{Figures 4c and 4d} show the CdSe control. Both the diagonal and
anti-diagonal projections are essentially stationary on this timescale.
The relevant excitonic structure is already encoded in the static
quantum dot Hamiltonian, and lattice motion only modulates this
pre-existing spectrum. The contrast with CsPbBr\({}_{3}\) rules out
projection geometry, finite bandwidth, or spectral congestion as the
origin of the delayed AD splitting. The effect is material-specific and
dynamical: in perovskite quantum dots, strong exciton--lattice coupling
separates the formation of excitonic states from the later formation of
coherent coupling between them.

The central implication of \textbf{Figures 2-4} is that the excitonic Hamiltonian
of the perovskite quantum dot is not fully present at the moment of
optical excitation. The optical field prepares \(X_{1}\), but
\(X_{2}\) appears only after the collective polaron field forms. The
accessible Hilbert space therefore expands in time, from
\(\left\{ \mid 0\rangle, \mid X_{1}\rangle \right\}\)to
\(\left\{ \mid 0\rangle, \mid X_{1}\rangle, \mid X_{2}\rangle \right\}\).
Once this manifold exists, the diagonal splitting becomes nearly static,
while the anti-diagonal splitting continues to grow as coherent
\(X_{1} - X_{2}\) coupling matures. CdSe shows none of this behavior. The
perovskite response therefore demands a state-generating lattice
coordinate: a collective polaron order parameter that creates the
excitonic manifold and subsequently couples its emergent states.

\vspace{-0.5cm}
\subsection{3. Lattice spectral density and polaron-field formation}

\textbf{Figure 5} shows that the decisive difference between CdSe and the
perovskite quantum dots is not merely the phonon frequency, but the
entire character and strength of the bath. CdSe is the weak-coupling
limit. Its Raman response \cite{ref63,ref78,ref79,ref80} is dominated by a
well-resolved LO phonon near \(26\ meV\), with only weak exciton--phonon
coupling, \(S \sim 0.03\). The lattice therefore acts as a perturbative
bath: it can modulate, broaden, or dephase an exciton, but it does not
reorganize the excitonic Hilbert space. This is why CdSe shows a
pre-existing D and AD response at time zero. The excitonic manifold is
already present, and the phonon bath only imposes small oscillatory
corrections on it.

\begin{figure}[tbp]
\centering
\includegraphics[width=0.95\linewidth]{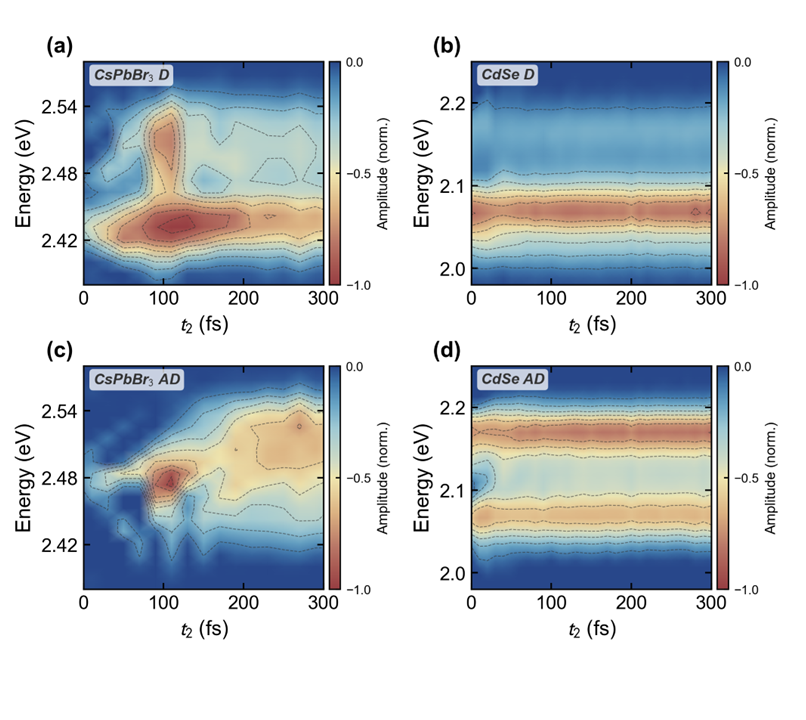}
\caption{Transient diagonal and anti-diagonal projections in perovskite and CdSe quantum dots. (a) Time-resolved diagonal projection for CsPbBr\({}_{3}\). The D response appears after excitation and rapidly develops a split structure, indicating formation of the polaron-generated two-state energy landscape. (b) Time-resolved diagonal projection for CdSe. The D response is present essentially at time zero and remains nearly stationary, consistent with a pre-existing excitonic manifold. (c) Time-resolved anti-diagonal projection for CsPbBr\({}_{3}\). The AD response is weak at early \(t_{2}\), then grows into a resolved split structure on the polaron-formation/coherence-onset timescale, revealing delayed coherent \(X_{1}\)--\(X_{2}\) coupling. (d) Time-resolved anti-diagonal projection for CdSe. The AD response is already present and shows little delayed growth, as expected for static excitonic coupling modulated by conventional phonons. Together, the projections show that CsPbBr\({}_{3}\) separates state formation from coherent coupling, whereas CdSe behaves as a conventional quantum dot with a pre-existing excitonic Hamiltonian.}
\label{fig:4}
\end{figure}

In contrast, the perovskite quantum dots have \(S \sim 1\) and do not
show a single dominant, weakly coupled LO mode. They exhibit a broad
low-frequency spectral density extending toward zero energy, the
signature of a soft, overdamped, strongly polar lattice. This is a
qualitatively different system--bath problem. The bath is no longer a
weak external reservoir that destroys coherence after the fact; it is
strongly coupled to the electronic excitation and reorganizes in
response to it. That reorganization is the quantum back-action of the
exciton on the lattice, and the back-action of the reorganized lattice
on the exciton.
\begin{figure}[tbp]
\centering
\includegraphics[width=0.92\linewidth]{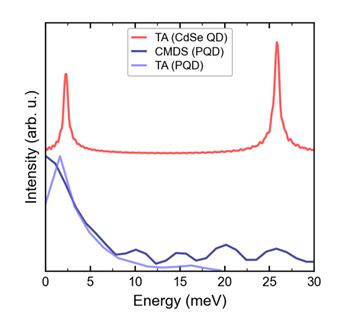}
\caption{Low-frequency lattice spectral density of perovskite and CdSe nanocrystals. \\
Low-frequency vibrational spectra extracted from transient absorption and CMDS measurements. CdSe quantum dots exhibit discrete, underdamped phonon modes, characteristic of a rigid semiconductor quantum dot. In contrast, the perovskite quantum-dot response is dominated by broad low-frequency spectral weight extending toward zero energy. This overdamped spectral density provides the collective lattice coordinate required for diffusive polaron formation and accounts for the \(\tau_{p}\)-scale spectral-diffusion dynamics observed in the CMDS linewidth response.}
\label{fig:5}
\end{figure}

\vspace{-0.1 cm}
The result is a self-consistent exciton--polaron field. In this regime,
strong system--bath coupling does not simply decohere the system. It
creates the effective potential, expands the optically accessible
Hilbert space, generates the \(X_{1}/X_{2}\) manifold, and only then
allows coherent \(X_{1}\)-\(X_{2}\) coupling to emerge. \textbf{Figure 5}
therefore supplies the microscopic origin of the central result: the
many-body excitonic states are born from strong coupling to a
structured, low-frequency polar bath, not protected by isolation from
it.

To describe this situation microscopically, the lattice cannot be
introduced merely as a bath that dephases excitonic states after those
states have already been defined. It must enter as a polar coordinate
capable of changing the excitonic basis itself. We therefore begin from
a Fröhlich-type exciton--phonon Hamiltonian,

\begin{equation}
\begin{aligned}
H =&\;
\sum_{n}\epsilon_n^{0}\lvert n\rangle\langle n\rvert
+\sum_{\mathbf q}\hbar\omega_{\mathbf q}
\,b_{\mathbf q}^{\dagger}b_{\mathbf q}
\\[3pt]
&+
\sum_{mn,\mathbf q}
g_{mn}(\mathbf q)\,
\lvert m\rangle\langle n\rvert
\left(
b_{\mathbf q}^{\dagger}
+b_{-\mathbf q}
\right).
\end{aligned}
\label{eq:Hamiltonian}
\end{equation}
Here \(\mid n\rangle\) are bare excitonic configurations of the
confined quantum dot, \(b_{\mathbf{q}}^{\dagger}\) creates a polar
lattice excitation, and \(g_{mn}(\mathbf{q})\) is the long-range
exciton--phonon coupling. The significance of Eq. 1 is that the lattice
couples not only to exciton energies, but to the structure of the
excitonic manifold. The diagonal terms \(g_{nn}\) generate the
polarization field associated with a given excitonic charge
distribution, while the off-diagonal terms \(g_{mn}\) allow that field to
mix excitonic configurations. In CdSe, these couplings remain
perturbative: they modulate a pre-existing manifold through relatively
weak normal-mode motion. In lead-halide perovskites, the soft polar
lattice makes the same interaction collective and non-perturbative. The
lattice polarization does not merely dress the exciton; it helps create
the exciton--polaron manifold whose splittings are measured
experimentally.

The physical object that forms after excitation is the coherent
lattice-polarization field,

\[\begin{matrix}
 & P(\mathbf{r},t) = \sum_{\mathbf{q}}^{}P_{\mathbf{q}}\langle b_{\mathbf{q}}^{\dagger}(t) + b_{- \mathbf{q}}(t)\rangle e^{i\mathbf{q} \cdot \mathbf{r}}. & & \text{(2)}
\end{matrix}
\]This field is the order parameter for polaron formation. Before
photoexcitation, the optically selected ensemble has no exciton-induced
polarization field. After excitation, the exciton acts as a source for
the polar lattice, and \(P(\mathbf{r},t)\) grows toward a self-consistent
configuration. The polaron is therefore not a static label attached to a
bare exciton. It is the ordered state of the exciton-coupled lattice
polarization field.

For the optical response, the full polarization field can be
coarse-grained into the collective coordinate most strongly coupled to
the exciton,

\[\begin{matrix}
 & Q(t) = \int d^{3}r\text{\,}W_{X}(\mathbf{r})P(\mathbf{r},t), & & \text{(3)}
\end{matrix}
\]where \(W_{X}(\mathbf{r})\) contains the exciton charge distribution
and the Fröhlich interaction kernel. Equivalently, \(Q(t)\) is the
collective component of the phonon displacement field selected by the
exciton. The Raman-derived low-frequency spectral density constrains the
dynamics of this coordinate. In perovskite quantum dots, this response
is overdamped and relaxational rather than a long-lived coherent normal
mode, so the formation of the polaron field can be represented by

\[\begin{matrix}
 & Q(t) = Q_{\infty}\left\lbrack 1-e^{- (t - t_{0})/\tau_{p}} \right\rbrack\Theta(t - t_{0}). & & \text{(4)}
\end{matrix}
\]Here \(t_{0}\) is the onset time for formation of the exciton-induced
polarization field and \(\tau_{p}\) is the relaxation time of the
collective polaron coordinate. The AD-linewidth dynamics show that this
polaron formation is largely complete by \(\sim 200\) fs. This timescale
matches the onset of the D and AD peak structures, showing that both
require formation of the polaron-renormalized manifold.

Once \(Q(t)\) becomes finite, the optically accessed Hilbert space is no
longer limited to the ground state and the initially prepared exciton.
At early time, the basis is effectively
\(\left\{ \mid 0\rangle, \mid X_{1}\rangle \right\}\). After the
collective polarization field forms, a second excitonic configuration is
stabilized and the basis expands to
\(\left\{ \mid 0\rangle, \mid X_{1}\rangle, \mid X_{2}\rangle \right\}\).
The appearance of \(X_{2}\) after \(\sim 100\)fs is therefore not
relaxation into a pre-existing level. It is the formation of a new
member of the polaron-renormalized excitonic manifold. In the
\(\left\{ \mid X_{1}\rangle, \mid X_{2}\rangle \right\}\) subspace, the
instantaneous polaron-frame Hamiltonian is

\[\begin{matrix}
 & H_{eff}\lbrack Q(t)\rbrack = \begin{pmatrix}
E_{1}\lbrack Q(t)\rbrack & V_{12}\lbrack Q(t)\rbrack \\
V_{12}^{*}\lbrack Q(t)\rbrack & E_{2}\lbrack Q(t)\rbrack
\end{pmatrix}. & & \text{(5)}
\end{matrix}
\]The diagonal and anti-diagonal splittings are different matrix
elements of this Hamiltonian:
\(\Delta_{D}(t) \simeq E_{2}\lbrack Q(t)\rbrack - E_{1}\lbrack Q(t)\rbrack\),
whereas \(\Delta_{AD}(t) \simeq 2 \mid V_{12}\lbrack Q(t)\rbrack \mid\).
This is the central identification. The D splitting measures formation
of the polaron-generated energy landscape. The AD splitting measures
coherent coupling between states inside that landscape. They are not two
equivalent projections of the same static splitting.

The delayed appearance of \(X_{2}\) explains why both D and AD structures
have an onset near the polaron-formation time. Before \(Q(t)\) forms,
there is no resolved two-state exciton--polaron manifold and therefore
no meaningful \(X_{1} - X_{2}\) splitting to measure. Once the lattice
polarization stabilizes \(X_{2}\), the diagonal separation
\(E_{2} - E_{1}\) becomes visible and then changes only weakly; this is
why the D splitting appears static after its onset. The AD splitting has
the same onset because coherent coupling cannot exist before both
\(X_{1}\) and \(X_{2}\) are present. Its magnitude, however, continues to
grow after linewidth-defined polaron formation is mostly complete. Thus
the AD splitting is not simply a clock for polaron formation. It is a
later stage: the maturation of coherent off-diagonal coupling inside an
already formed exciton--polaron manifold.

We express this separation by writing the coupling as

\[\begin{matrix}
 & V_{12}(t) = \lambda_{12}Q(t)S_{12}\lbrack Q(t)\rbrack C_{12}(t). & & \text{(6)}
\end{matrix}
\]Here \(\lambda_{12}\) is the microscopic off-diagonal exciton--phonon
coupling, \(S_{12}\lbrack Q(t)\rbrack\) is the overlap between the two
polaron-dressed configurations, and \(C_{12}(t)\) is the
phase-correlation factor for coherent inter-state coupling. The
collective coordinate \(Q(t)\) creates the polaronic manifold and enables
\(X_{2}\). The factor \(C_{12}(t)\) describes the later development of
coherent coupling within that manifold. Thus the AD splitting naturally
shares the onset of the D splitting but grows over a longer interval.

The smaller magnitude of the AD splitting follows from the same
structure. In CsPbBr\({}_{3}\), the diagonal splitting is the full
energy separation generated by the polaron-renormalized potential, while
the anti-diagonal splitting is only the coherent off-diagonal projection
of that potential. In compact form,
\(\Delta_{D} \sim \mid E_{2} - E_{1} \mid\), but
\(\Delta_{AD} \sim 2 \mid V_{12} \mid\). Because \(V_{12}\) is controlled
by overlap and phase coherence between distinct polaron-dressed
configurations, it is naturally smaller than the full diagonal energy
separation. The hierarchy \(\Delta_{AD} < \Delta_{D}\) is therefore not a
generic property of any two-state spectrum. It is a signature of the
perovskite exciton--polaron manifold, where state formation and coherent
coupling are distinct physical processes.

CdSe provides the limiting contrast. There, the excitonic states and
their couplings are already present in the static quantum dot
Hamiltonian \cite{ref11,ref12,ref13,ref81}, and the lattice response is
better described as a perturbative modulation by discrete normal modes,
\(H_{CdSe}(t) \simeq H_{ex}^{0} + \delta H_{ph}(t)\). In this case, the
lattice does not create a new manifold during \(t_{2}\). D and AD
projections therefore report the same pre-existing coupled-state
structure, with no delayed AD growth and no perovskite-like reduction of
AD relative to D. This difference is the decisive control: the D/AD
separation is not a projection artifact, but a consequence of dynamical
polaron formation in the perovskite lattice.

\subsection{4. Density-matrix model of delayed coherence formation}

The density-matrix formulation makes the Hilbert-space expansion
explicit. Before polaron formation, the optically active density matrix
is effectively two-dimensional,

\[\begin{matrix}
 & \rho_{0X_{1}}(t) = \begin{pmatrix}
\rho_{00}(t) & \rho_{01}(t) \\
\rho_{10}(t) & \rho_{11}(t)
\end{pmatrix}, & & \text{(7)}
\end{matrix}
\]corresponding to the basis
\(\left\{ \mid 0\rangle, \mid X_{1}\rangle \right\}\). After \(X_{2}\) is
stabilized by the polaron field, the relevant density matrix expands to

\[\begin{matrix}
 & \rho(t) = \begin{pmatrix}
\rho_{00}(t) & \rho_{01}(t) & \rho_{02}(t) \\
\rho_{10}(t) & \rho_{11}(t) & \rho_{12}(t) \\
\rho_{20}(t) & \rho_{21}(t) & \rho_{22}(t)
\end{pmatrix}. & & \text{(8)}
\end{matrix}
\]The new diagonal element \(\rho_{22}\) corresponds to spectral weight
in the newly formed \(X_{2}\) configuration, while \(\rho_{12}\) is the
inter-excitonic coherence that produces the AD response. Its evolution
is governed by

\[\begin{matrix}
 & \frac{d\rho_{12}}{dt} = - \left\lbrack i\omega_{12}(t) + \gamma_{12} \right\rbrack\rho_{12} - \frac{i}{\hslash}V_{12}(t)\left\lbrack \rho_{22}(t) - \rho_{11}(t) \right\rbrack, & & \text{(9)}
\end{matrix}
\]with
\(\hslash\omega_{12}(t) = E_{2}\lbrack Q(t)\rbrack - E_{1}\lbrack Q(t)\rbrack\).
This equation shows why a finite D splitting does not imply an immediate
or equal AD splitting. The diagonal response requires the appearance of
two resolved energies, \(X_{1}\) and \(X_{2}\). The anti-diagonal
response requires the off-diagonal density-matrix element \(\rho_{12}\),
which is driven by the time-dependent coupling \(V_{12}(t)\). In
CsPbBr\({}_{3}\), the polaron field first expands the Hilbert space and
generates the two-state landscape; only later does it drive the coherent
off-diagonal response. In CdSe, the equivalent states and couplings are
already contained in \(H_{ex}^{0}\), so D and AD appear together without
the dynamical separation observed in the perovskite quantum dots.

The theory therefore gives a direct interpretation of the dynamics in
\textbf{Figure 4}. The IRF-limited response corresponds to optical preparation of
\(X_{1}\). The growth of the AD linewidth and fractional
\(X_{2}\) amplitude follows the polaron coordinate \(Q(t)\), identifying
\(\tau_{p}\) as the formation time of the collective lattice-polarization
field. Once \(Q(t)\) is finite, the excitonic Hilbert space expands from
the optically prepared \(X_{1}\) configuration to the
\(X_{1}/X_{2}\) manifold, producing the diagonal splitting. The
anti-diagonal splitting appears later because it requires the
off-diagonal density-matrix element \(\rho_{12}\), not merely the
existence of two states. Its delayed growth therefore reflects coherence
formation within an already formed exciton-polaron manifold, governed
by the distributed onset of \(X_{1}/X_{2}\) coupling and the intrinsic
buildup time \(\tau_{c}\). Thus Fig. 4 is not a set of unrelated kinetic
traces: it is the time-domain projection of a single sequence-optical
preparation, polaron-field formation, Hilbert-space expansion, and
delayed coherent coupling.

\textbf{Figure 6} brings the dynamical and scaling results into a single forward
model. The first question is whether the polaron formation rate and the
polaron-induced energy scale are independent empirical quantities, or
whether both are controlled by the same low-frequency lattice spectral
density. \textbf{Figure 6a} shows that they are linked. For CsPbBr\({}_{3}\), the
measured size series defines a nearly linear branch relating the polaron
formation rate \(k_{polaron}\) to the spectral-diffusion amplitude
\(\Delta\Gamma_{SD}\). This correlation is not a trivial consequence of
linewidth fitting. The horizontal axis measures the magnitude of the
low-frequency energy-gap fluctuations generated by the polar lattice,
whereas the vertical axis measures the rate at which those fluctuations
reorganize into the ordered exciton-polaron state. Their
proportionality shows that, within a fixed composition, the same
overdamped lattice coordinate controls both the fluctuation amplitude
and the formation kinetics.

\begin{figure}[tbp]
\centering
\includegraphics[width=0.92\linewidth]{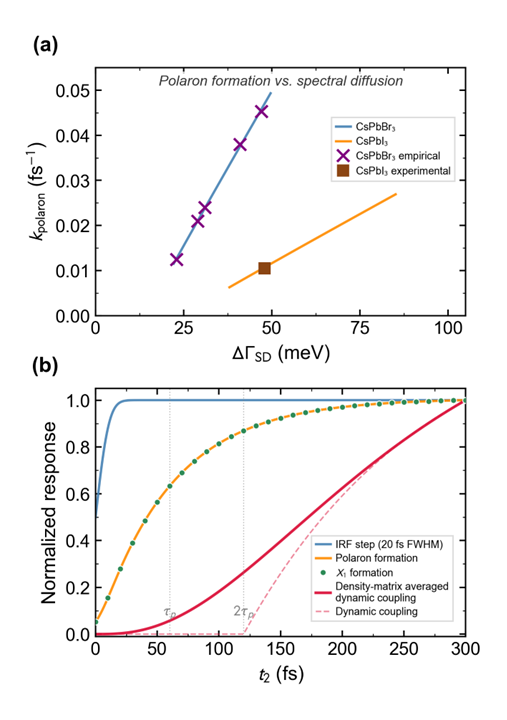}
\caption{Polaron formation, spectral diffusion, and density-matrix forward model. (a) Correlation between the polaron formation rate \(k_{polaron}\) and the polaron-induced spectral-diffusion amplitude \(\Delta\Gamma_{SD}\). CsPbBr\({}_{3}\) defines a steep kinetic branch: larger spectral diffusion is associated with faster polaron formation. CsPbI\({}_{3}\) lies on a lower-slope iodide branch, showing that the softer iodide lattice produces larger fluctuation amplitudes but slower reorganization. (b) Forward model for the temporal observables. The \(X_{1}\) response follows the 20 fs IRF, while polaron formation and fractional \(X_{2}\) appearance follow the same \(\tau_{p}\)-limited rise. The AD splitting is delayed until the \(X_{1}/X_{2}\) manifold exists and then grows as the density-matrix-averaged coherence \(\langle \mid \rho_{12}(t) \mid \rangle\), with dynamic coupling beginning near \(2\tau_{p}\).}
\label{fig:6}
\end{figure}

\textbf{Figure 6a} extends the relation between polaron formation and spectral
diffusion from CsPbBr\({}_{3}\) to CsPbI\({}_{3}\). For
CsPbBr\({}_{3}\), the measured size series defines a steep branch
relating the polaron formation rate \(k_{polaron}\) to the
spectral-diffusion amplitude \(\Delta\Gamma_{SD}\). The horizontal axis
measures the magnitude of the low-frequency energy-gap fluctuations
produced by the polar lattice, whereas the vertical axis measures the
rate at which those fluctuations reorganize into the exciton-polaron
field. Their near-linear relation shows that, within the bromide
lattice, the same overdamped lattice coordinate controls both
fluctuation amplitude and polaron formation kinetics.

Replacing Br by I shifts the system onto a different branch.
CsPbI\({}_{3}\) exhibits a larger polaron-induced energy scale,
consistent with the greater softness and polarizability of the iodide
lattice, but the slope is reduced. Thus the iodide lattice generates a
deeper polarization field while forming it more slowly. The
CsPbI\({}_{3}\) experimental point falls on this lower-slope iodide
branch, showing that the strength of the polaron field and the rate at
which it forms are linked but not controlled by a single scalar coupling
constant. The amplitude reflects the reorganization strength of the soft
polar lattice; the rate reflects the bandwidth and damping of the
low-frequency spectral density. \textbf{Figure 6a} therefore separates two
aspects of the same quantum back-action: how strongly the lattice
polarizes in response to the exciton, and how quickly that polarization
field becomes the ordered exciton--polaron coordinate.

\textbf{Figure 6b} converts this phenomenology into a density-matrix forward
model for the time-domain observables. The optical field first prepares
\(X_{1}\) within the 20 fs instrument response. The polaron coordinate
\(Q(t)\) then grows with \(\tau_{p}\), producing both the measured
\(AD_{X_{1}}\) linewidth increase and the fractional \(X_{2}\) appearance.
These two observables are therefore the same clock: they measure
formation of the collective lattice-polarization field and the
associated expansion of the optically accessible excitonic manifold. The
AD splitting is different. It cannot appear until the
\(X_{1}/X_{2}\) manifold exists, and it is therefore delayed relative to
both optical preparation and polaron formation.

In the density-matrix picture, the AD splitting is proportional to the
ensemble-averaged off-diagonal element
\(\langle \mid \rho_{12}(t) \mid \rangle\). A single quantum dot
develops this coherence only after its local polaron field has generated
the two-state manifold. The ensemble response is therefore an average
over polaron-formation onset times distributed around \(2\tau_{p}\),
followed by an intrinsic coherence-building time \(\tau_{c}\). This
construction explains why the AD splitting begins only after the D and
AD peak structures appear, and why it continues to grow after the AD
linewidth dynamics have nearly saturated. The polaron field creates the
states; the off-diagonal density-matrix element forms later.

Together, the two panels establish the central dynamical structure. The
low-frequency lattice spectral density controls \(\tau_{p}\),
\(\Delta\Gamma_{SD}\), and the composition-dependent phase diagram of
polaron formation. The later AD splitting reflects a higher-order
process: coherent coupling inside the already formed exciton--polaron
manifold. Thus the perovskite response is not a static excitonic
spectrum broadened by a bath. It is a sequence of many-body events:
optical preparation, collective lattice reorganization, Hilbert-space
expansion, and delayed coherence formation.

\subsection{5. Size, composition, and radiative-rate scaling}

\textbf{Figure 7} tests the origin of the diagonal and anti-diagonal splittings
by asking how they scale with nanocrystal edge length. The result is
incompatible with ordinary quantum confinement \cite{ref11,ref81}.
Both splittings increase approximately linearly with edge length,
whereas single-particle level spacings must decrease as the dot becomes
larger. The splittings are therefore not 1P--1S gaps, excitonic
splittings, or static confinement energies. They are collective
exciton--lattice energy scales.

\begin{figure*}[tbp]
\centering
\includegraphics[width=0.95\textwidth]{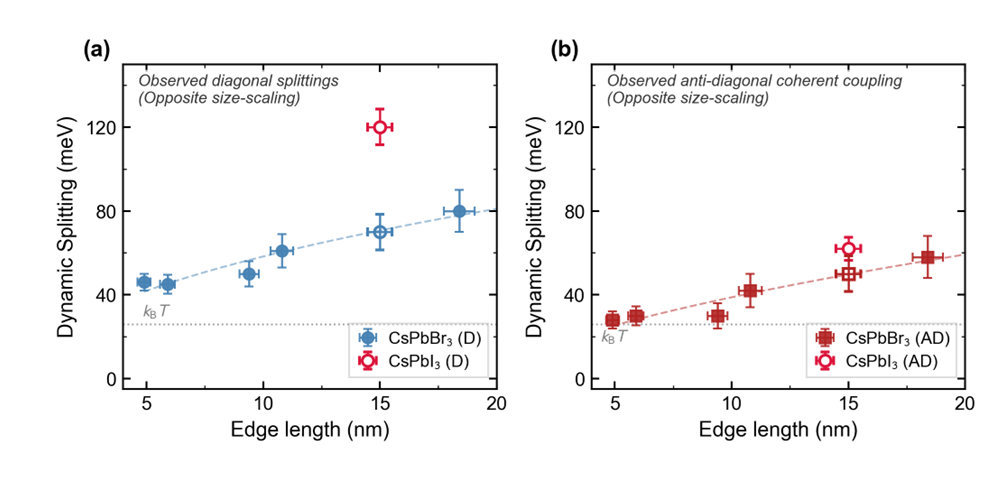}
\caption{Opposite size scaling of diagonal and anti-diagonal exciton--polaron splittings. (a) Diagonal splitting \(D\) as a function of quantum-dot edge length. In CsPbBr\({}_{3}\), \(D\) increases with size, opposite to the decrease expected for ordinary quantum confinement. The CsPbI\({}_{3}\) point lies at substantially larger splitting, consistent with stronger polaronic stabilization in the softer iodide lattice. (b) Anti-diagonal splitting \(AD\) as a function of edge length. The \(AD_{splitting}\) also increases with size, showing that coherent \(X_{1}\)-\(X_{2}\) coupling grows with the spatial extent of the exciton--polaron manifold rather than behaving as a fixed fine-structure or vibronic splitting. The smaller magnitude of \(AD\) relative to \(D\) shows that only part of the polaron-generated energy landscape appears as coherent off-diagonal coupling. The dotted line marks \(k_{B}T\) at room temperature.}
\label{fig:7}
\end{figure*}

The D splitting is systematically larger than the AD splitting at every
size, but the two trends are parallel. This is the crucial observation.
The AD splitting is not governed by a different size law; it is a
reduced projection of the same growing polaronic energy scale. The D
splitting measures the full energetic separation between two
polaron-dressed excitonic configurations, whereas the AD splitting
measures the coherent off-diagonal coupling between them. Thus both
splittings grow with the spatial extent of the same exciton-induced
polarization field, but only part of that field appears as
phase-coherent \(X_{1} - X_{2}\) mixing.

\begin{figure*}[tbp]
\centering
\includegraphics[width=0.95\textwidth]{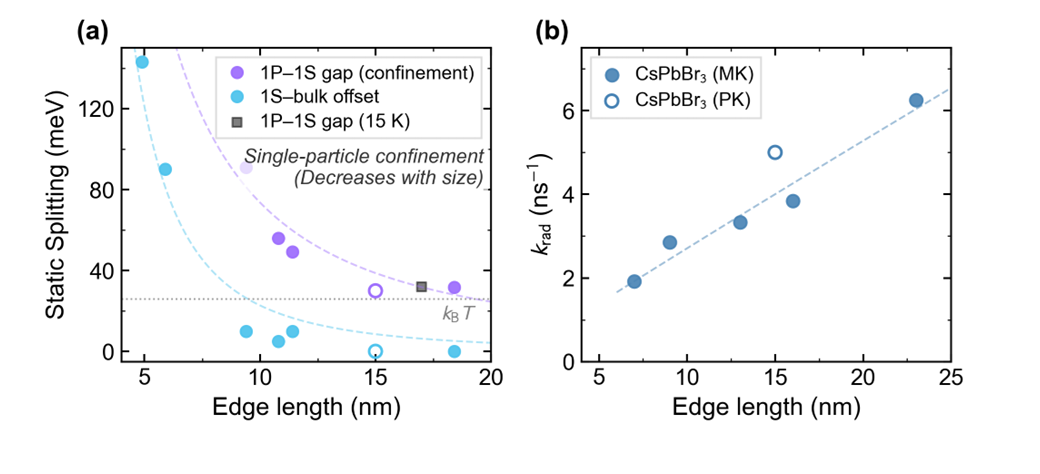}
\caption{Comparison with single-particle confinement and radiative-rate scaling. (a) The observed D and AD splittings are compared with conventional single-particle confinement energies, including the 1P-1S gap and the 1S-to-bulk offset. Unlike ordinary quantum-confinement energies, which decrease with increasing size, the measured splittings show the opposite trend, demonstrating that they do not arise from simple single-particle level spacing. (b) The extracted splitting is plotted against quantum-dot edge length together with the radiative rate, \(k_{r}\), measured independently. Both exhibit the same approximately linear size dependence, suggesting that the emergent coherent coupling and enhanced radiative response share a common origin in exciton-polaron formation.}
\label{fig:8}
\end{figure*}

The CsPbI\({}_{3}\) point exposes the difference between diagonal
stabilization and coherent coupling. Iodide produces a much larger D
splitting but only a moderately larger AD splitting. The softer, more
polarizable iodide lattice therefore deepens the polaronic energy
landscape more efficiently than it increases coherent inter-state
mixing. In the D/AD representation, CsPbI\({}_{3}\) forms a rectangular
rather than square exciton-polaron box: the diagonal energy separation
is strongly enhanced, while the anti-diagonal coupling grows only
modestly. The comparison with \(k_{B}T\) at 300 K further shows that
these are not small perturbative corrections. The splittings are
room-temperature-scale energy splittings of an emergent many-body
manifold.

This size dependence follows naturally if the relevant energy is
generated by a long-range Fröhlich polarization field. The
polaron-induced contribution to the excitonic Hamiltonian is controlled
by the interaction of the excitonic charge density with the
self-consistent lattice-polarization potential,

\[\begin{matrix}
 & \Delta_{pol}(L) \sim \int_{V_{L}}^{}d^{3}r\text{\,}\rho_{X}(\mathbf{r})\Phi_{P}(\mathbf{r}). & & \text{(10)}
\end{matrix}
\]Here \(\rho_{X}(\mathbf{r})\) is the excitonic charge-density
difference, \(\Phi_{P}(\mathbf{r})\) is the electrostatic potential
produced by the ordered polarization field, and \(L\) is the quantum-dot
edge length. A short-range deformation-potential or particle-in-a-box
effect would not naturally grow with \(L\). A Fröhlich field is
different: it is long-ranged, and the coherent polarization length can
be limited by the quantum dot itself. Over the measured range, this
gives the empirical scaling \(\Delta_{pol}(L) = aL + \Delta_{0}\). The
observed ``wrong-way'' size dependence is therefore the signature of a
growing collective polarization field, not a growing confinement energy.

The diagonal and anti-diagonal splittings are then two projections of
this same field. For CsPbBr\({}_{3}\), the data are captured by
\(\Delta_{D}(L) = aL + \Delta_{D}^{0}\) and
\(\Delta_{AD}(L) = aL + \Delta_{AD}^{0}\). The common slope reflects the
common spatial origin. The offset reflects the fact that AD measures
coherent coupling rather than the full diagonal energy separation.
Equivalently, \(\Delta_{D}(L) - \Delta_{AD}(L)\) is approximately size
independent over the measured range. This offset is the energetic cost
of projecting the polaron-generated landscape onto an off-diagonal
coherence. D reports the energy landscape; AD reports the coherent
mixing inside it.

Composition changes the same Hamiltonian anisotropically. Replacing Br
by I increases lattice polarizability and softens the polar lattice,
producing stronger diagonal polaron stabilization. Thus CsPbI\({}_{3}\)
develops a much larger D splitting. The AD splitting does not increase
proportionally because \(V_{12}\)contains not only polarization strength
but also wavefunction overlap and phase correlation; in compact form,
\(V_{12}(L) \sim \alpha_{F}LS_{12}C_{12}\). A softer iodide lattice can
make the two polaron-dressed configurations more separated while not
equivalently increasing their coherent overlap. This is why iodide
produces a rectangular exciton--polaron box rather than a uniformly
scaled one.

\textbf{Figure 8} places this anomalous size scaling beside ordinary quantum-size
effects and radiative-rate constant data at 10 K showing superradiance
\cite{ref46,ref47}. \textbf{Figure 8a} shows the expected behavior for
single-particle confinement: the 1P-1S gap and the 1S-to-bulk offset
both decrease with edge length as the dot approaches the bulk limit. The
D and AD splittings do the opposite. \textbf{Figure 8b} then shows that the
exciton radiative rate constant, \(k_{r}\), follows the same increasing
edge-length dependence as the splittings. This comparison is powerful
because \(k_{r}\) is measured by time-resolved photoluminescence, whereas
D and AD are measured by CMDS. Different experiments, different
observables, yielding the same length scaling.

The common scaling points to a common physical object: a spatially
extended exciton--polaron coherence limited by the quantum dot edge
length. CMDS measures its Hamiltonian matrix elements: D is the diagonal
energy separation and AD is the off-diagonal coherent coupling. TRPL
measures its oscillator strength. In the collective-emission framework,
\(k_{r}\)grows because the effective bright-state dipole increases with
the number of phase-locked unit-cell dipoles participating in the
exciton--polaron state; equivalently,
\(k_{r}^{X}(T,D) \propto N_{eff}(T,D) \mid \mu_{0} \mid^{2}\), where
\(N_{eff}\) is the coherence participation number. Over the measured
range, the data imply \(N_{eff}(L) \propto L\), giving the same linear
edge-length law seen in \(\Delta_{D}\) and \(\Delta_{AD}\). The emergent
excitonic structure and the enhanced radiative response are therefore
not separate anomalies. They are two measurements of the same collective
exciton--lattice state.

\subsection{6. Conclusions}

The central result of this work is that, in a soft polar semiconductor,
the excitonic spectrum is not simply revealed by light; it is assembled
in time by quantum back-action between electronic polarization and
lattice polarization. CdSe gives the conventional limit: the optical
pulse interrogates confined excitons whose Hamiltonian is already
present, while a weak, discrete phonon bath perturbs that spectrum. The
perovskite lies in the opposite limit. The pulse first prepares an
\(X_{1}\) polarization; the polar lattice then responds, forming a
collective field that expands the optically accessible excitonic Hilbert
space so that \(X_{2}\) enters the manifold; coherent
\(X_{1}\)-\(X_{2}\) coupling appears only later. The experiment
therefore resolves quasiparticle birth as a sequence of events---optical
preparation, lattice-field formation, Hilbert-space expansion, and
delayed coherence---rather than as a static spectrum dressed after the
fact. The Raman response identifies the origin of this hierarchy: the
perovskite is strongly coupled to a broad low-frequency spectral density
with order-unity Huang--Rhys strength, not weakly coupled to a single
perturbative phonon. The finite-crystal scaling further shows that the
relevant coordinate is not quantum confinement but an extended
polarization field of the soft lattice: the diagonal and anti-diagonal
splittings strengthen with polarizable volume and track the growth of
radiative oscillator strength. CMDS and TRPL therefore observe different
projections of the same exciton--polaron object, one through Hamiltonian
matrix elements and the other through optical strength. The implication
is broader than perovskite nanocrystals: strong system--bath coupling
can be constructive. In this regime the environment is not merely a
reservoir that dephases quantum states; it is the dynamical field that
expands the manifold in which those states exist and later couples them
coherently. Polaron formation is therefore not a relaxation channel
following excitation, but a route to engineering many-body optical
structure in soft quantum matter.

\vspace{-0.1cm}
\subsection*{Supplementary Information}

Details of materials synthesis and characterization, coherent
multi-dimensional spectroscopy, theory.

\subsection{Acknowledgements}

P.K. acknowledges financial support from CFI, NSERC, and Sony. M.K.
acknowledges financial support from European Research Council through
the European Union's Horizon 2020 programme (ERC Consolidator Grant
SCALE-HALO, agreement no. 819740).

\end{document}